\documentclass[final,1p,times]{elsarticle}

\usepackage[normalem]{ulem}
\usepackage{amsmath,amssymb,xspace,amsthm,ctable,url,multirow,wrapfig,txfonts,fge}
\usepackage{pstricks,pst-node,pst-plot,pst-coil,pst-text,multido,pst-3d,pst-tree,pst-grad}
\usepackage{trfsigns,graphics,pspicture,nicefrac,longtable} 
\usepackage{trfsigns,graphics,pspicture,nicefrac,longtable}

\biboptions{sort&compress}

\journal{Discrete Applied Mathematics}

\newtheorem{theorem}{Theorem}[section]

\newtheorem{claim}{Claim}[theorem]

\newcommand{\EG}{{\it e.g.}\xspace}
\newcommand{\IE}{{\em i.e.}\xspace}
\newcommand{\tx}{^{\rm th}}

\newcommand{\NP}{\mbox{NP}}

\newcommand{\eps}{\varepsilon}

\newcommand{\comment}[1]{}

\begin{document}

\begin{frontmatter}

\title{On optimal approximability results for computing the strong metric dimension\tnoteref{label1}}

\tnotetext[label1]{Research partially supported by NSF grants IIS-1160995.}

\author[add]{Bhaskar DasGupta\corref{cor1}}
\ead{bdasgup@uic.edu}
\ead[url]{http://www.cs.uic.edu/~dasgupta}

\author[add]{Nasim Mobasheri}
\ead{nmobas2@uic.edu}

\cortext[cor1]{Corresponding author.}

\address[add]{Department of Computer Science, University of Illinois at Chicago, Chicago, IL 60607, USA}

\begin{abstract}
The strong metric dimension of a graph was first introduced 
by Seb\"{o} and Tannier (Mathematics of Operations Research, 29(2), 383-393, 2004)
as an alternative to the 
(weak) metric dimension of graphs 
previously 
introduced independently by Slater (Proc. $6\tx$ Southeastern Conference on Combinatorics, Graph Theory, and Computing, 549-559, 1975)
and 
by Harary and Melter (Ars Combinatoria, 2, 191-195, 1976),
and has since been investigated in 
several research papers. However, the exact worst-case computational complexity of computing 
the strong metric dimension has remained open beyond being $\NP$-complete. 
In this communication, we show that the problem of computing 
the strong metric dimension of a graph of $n$ nodes 
admits a polynomial-time $2$-approximation, admits a $O^\ast\big(2^{\,0.287\,n}\big)$-time exact computation algorithm, 
admits a $O\big(1.2738^k+n\,k\big)$-time exact computation algorithm if  
the strong metric dimension is at most $k$, 
does not admit a polynomial time 
$(2-\eps)$-approximation algorithm
assuming the unique games conjecture is true, 
does not admit a polynomial time 
$(10\sqrt{5}-21-\eps)$-approximation algorithm
assuming P$\neq\NP$, 
does not admit a $O^\ast\big(2^{o(n)}\big)$-time exact computation algorithm
assuming the exponential time hypothesis is true, 
and 
does not admit a $O^\ast\big(n^{o(k)}\big)$-time exact computation algorithm
if the strong metric dimension is at most $k$
assuming the exponential time hypothesis is true.
\end{abstract}


\end{frontmatter}

\section{Introduction}

\newcommand{\cP}{\mathcal{P}}
\newcommand{\cA}{\mathcal{A}}
\newcommand{\Nbr}{\text{N}}
\newcommand{\diam}{\text{diam}}
\newcommand{\SMD}{{\sc Str-Met-Dim}\xspace}
\newcommand{\sdim}{\text{sdim}}
\newcommand{\binary}{\text{binary}}
\newcommand{\MNC}{\text{{\sc Mnc}\xspace}}
\newcommand{\UGC}{{\sc Ugc}\xspace}
\newcommand{\ETH}{{\sc Eth}\xspace}
\newcommand{\SAT}{{\sc Sat}\xspace}
\newcommand{\kSAT}{$k$-{\sc Sat}\xspace}
\newcommand{\tSAT}{$3$-{\sc Sat}\xspace}

The concept of the metric dimension of graphs was originally
introduced independently by Slater~\cite{S75}
and 
by Harary and Melter~\cite{HM76} in the $1970$'s.
Their definition involved determining a minimum number of nodes such that distance vectors from each of these nodes
to all other nodes (the ``resolving vectors'') can be used to ''distinguish'' every pair of nodes in the graph.
Computing the metric dimension is known to be $\NP$-complete~\cite{GJ79}.
Optimal approximability results for the metric dimension was provided 
by Hauptmann \emph{et al.} 
in~\cite{HSV12} by showing both a 
$(\ln n+\ln \log_2 n + 1)$-approximation based on an approximation algorithm for test set 
problems in~\cite{BDK05} and also a $(1-\eps)$-inapproximability for any 
constant $0<\eps<1$.

Unfortunately, the metric dimension of a graph suffers from two difficulties, namely that 
the problem does not provably admit a better-than-logarithmic approximation and the resolving 
vectors cannot be used to uniquely identify the graph.
The \textbf{strong} metric dimension of a graph was therefore introduced by Seb\"{o} and Tannier~\cite{ST04} 
as an alternative to the above-mentioned metric dimension of graphs.
The resulting ``strongly'' resolving vectors can indeed be used to uniquely identify the given graph.
Subsequently, the strong metric dimension has been investigated in 
several research papers such as~\cite{OP07,RYKO14,Y13}. 
Let $G=(V,E)$ be a given undirected graph of $n$ nodes.
To define the strong metric dimension, we will use the
following notations and terminologies: 
\begin{itemize}
\item
${
\Nbr(u)}=\Big\{ \, v \,\big|\,\{u,v\}\in E \, \Big\}$ denotes the set of neighbors of a node $u$.
\item
$u\!\stackrel{{s}}{\leftrightsquigarrow}\! v$ denotes a shortest path from between nodes $u$ and $v$ of length (number of edges)
$d_{u,v}$.
\item
$\diam(G)=\max_{u,v\in V} \left\{ \,d_{u,v}\, \right\}$ denotes the diameter of a graph $G$.
\item
A shortest path 
$u\!\stackrel{{s}}{\leftrightsquigarrow}\! v$
is called \emph{maximal}\footnote{\comment{\color{blue}}The end-points of such a path is called a mutually maximally distant
pairs of nodes in~\cite{ST04}.}
if and only if it is not \emph{properly} included inside another shortest path, \IE, if and only if
the predicate
\[
\Big(
\,
\forall \, x\in\Nbr(u) \colon d(x,v)\leq d(u,v)
\,
\Big)
\,
\bigwedge
\,
\Big(
\,
\forall \, y\in\Nbr(v) \colon d(y,u)\leq d(u,v)
\,
\Big)
\]
is true.
\item
A node $x$ \emph{strongly resolves} a pair of nodes $u$ and $v$, denoted by $x\blacktriangleright \{u,v\}$, if and only if 
either $v$ is on a shortest path between $x$ and $u$, or $u$ is on a shortest path between $x$ and $v$.
\item
A set of nodes $V'\subseteq V$ is a \emph{strongly resolving set} for $G$, denoted by 
$V'\blacktriangleright G$, 
if and only if 
every distinct pair of nodes of $G$ is strongly resolved by some node in $V'$.
\end{itemize}
Then, the problem of computing the strong metric dimension of a graph can be defined as follows:

\begin{longtable}{r l}
\toprule
\textbf{Problem name}: & 
Strong Metric Dimension (\SMD)
\\
[4pt]
\textbf{Instance}: & 
an undirected graph $G=(V,E)$.
\\
[4pt]
\textbf{Valid Solution}: &
a set of nodes $V'\subseteq V$ such that $V'\blacktriangleright G$.
\\
[4pt]
\textbf{Objective}: & 
\emph{minimize} $|V'|$.
\\
[4pt]
\textbf{Related notation}: & 
$\displaystyle\sdim(G)=\min_{V'\subseteq V \,\, \wedge\,\,  V'\blacktriangleright\, G} \Big\{ \,\,\big|\, V'\, \big| \,\,\Big\}$.
\\
\bottomrule
\end{longtable}

\subsection{Standard Concepts From the Algorithms Research Community}

For the benefit of readers not familiar with analysis of approximation algorithms, we 
state below some standard definitions; see standard textbooks such as~\cite{GJ79,V01,FK10} for further details.
An algorithm for a minimization
problem is said to have an \emph{approximation ratio} of $\rho$ 
(or simply called a $\rho$-approximation) provided the algorithm runs in polynomial time in the size of the input and produces 
a solution with an objective value 
\emph{no larger than} $\rho$ times
the value of the optimum.
A computational problem $P$ is said to be $\rho$-inapproximable under a complexity-theoretic assumption of $\mathbb{A}$ 
provided, assuming $\mathbb{A}$ to be true, there exists no $\rho$-approximation for $P$.
The (standard) Boolean satisfiability problem when every clause has exactly $k$ literals
will be denoted by \kSAT.
Finally, for two functions $f(n)$ and $g(n)$ of $n$, we say 
$f(n)=O^\ast(g(n))$ if $f(n)=O(g(n)\,n^c)$ for some positive constant $c$.

\smallskip

\subsection{Brief Overview of Three Well-known Complexity Theoretic Assumptions}

For the benefit of those readers not well familiar with well-known 
c\-o\-m\-p\-l\-e\-x\-i\-t\-y-theoretic assumptions,
we provide a very brief overview of the three complexity-theoretic assumptions used in this communication.

\medskip
\noindent
{\bf The P$\neq\!\!\!\NP$ assumption}
Starting with the famous Cook's theorem~\cite{C71} in $1971$ and Karp's subsequent paper in $1972$~\cite{K72}, 
the P$\neq\!\!\!\NP$ assumption is 
the central assumption in structural complexity theory and algorithmic complexity analysis.

\medskip
\noindent
{\bf The Unique Games Conjecture (\UGC)}
The Unique Games Conjecture, formulated by Khot in~\cite{K02}, is one of the most 
important open question in computational complexity theory. Informally speaking,  
the conjecture states that,
assuming P$\neq\!\NP$,
a type of constraint satisfaction problems does not admit a polynomial time algorithm to distinguish 
between instances that are almost satisfiable from instances that are almost completely
unsatisfiable. There is a large body of research works showing that the conjecture has many interesting 
implications
and many researchers routinely assume \UGC to prove non-trivial inapproximability results.
An excellent survey on \UGC can be found in many places, for example in~\cite{T12}.

\medskip
\noindent
{\bf The Exponential Time Hypothesis (\ETH)}
In an attempt to provide a 
rigorous evidence that 
the complexity of \kSAT increases with increasing $k$,
Impagliazzo and Paturi in~\cite{IP01}
formulated the so-called
Exponential Time Hypothesis (\ETH) 
in the following manner.
Letting 
$s_k=\inf \big\{ \, \delta \,:\, \text{there exists } O^\ast\big(2^{\delta n}\big)$ algorithm for solving \kSAT$\big\}$, 
\ETH states that $s_k>0$ for all $k\geq 3$, \IE,
\kSAT does not admit a sub-exponential time (\IE, of time $O^\ast\big(2^{o(n)}\big)$) 
algorithm\footnote{For two functions $f(x)$ and $g(x)$ of $x$, $f=o(g)$ provided $\lim_{x\to\infty} f(x)/g(x)=0$.}.
\ETH has significant implications for worst-case time-complexity of exact solutions of search problems, \EG, 
see~\cite{IPZ01,W03}.

\subsection{Our Results}

Let $G=(V,E)$ be the given graph.
It is easy to see following the approach in Khuller \emph{et al.}~\cite{KRR96}
that the 
problem of computing the strong metric dimension 
$\sdim(G)$ 
can be reduced to an instance of the (unweighted) set-cover problem giving a $O(\log |V|)$-approximation.
In this communication, we show further improved results as 
summarized by the following theorem.

\begin{theorem}\label{main-thm}~\\
\noindent
{\bf (a)}
\SMD admits the following type of algorithms:
\begin{itemize}
\item
polynomial-time $2$-approximation, 
\item
$O^\ast\big(2^{\,0.287\,n}\big)$-time exact computation algorithm, and 
\item
$O\big(1.2738^k+n\,k\big)$-time exact computation algorithm where 
$\sdim(G)\leq k$. 
\end{itemize}

\noindent
{\bf (b)}
Assuming that the unique games conjecture
\emph{(} \UGC\emph{)}
is true, \SMD does not admit a polynomial-time $(2-\eps)$-approximation 
for any constant $0<\eps\leq 1$ 
even if the given graph 
is restricted in the sense that
\begin{quote}
\begin{description}
\item[(i)] 
$\diam(G)\leq 2$, or 
\item[(ii)] 
$G$ is bipartite and $\diam(G)\leq 4$.
\end{description}
\end{quote}

\medskip

\noindent
{\bf (c)}
Assuming \emph{P$\neq\!\!\NP$}, 
\SMD does not admit a polynomial-time 
$(10\sqrt{5}-21-\eps)$-approximation\footnote{Note that $10\sqrt{5}-21\approx 1.36068<2$.}
for any constant $0<\eps\leq 10\sqrt{5}-22$ 
even if the given graph 
is restricted in the sense that
\begin{quote}
\begin{description}
\item[(i)] 
$\diam(G)\leq 2$, or 
\item[(ii)] 
$G$ is bipartite and $\diam(G)\leq 4$.
\end{description}
\end{quote}
%

\medskip

\noindent
{\bf (d)}
Assuming the exponential time hypothesis (\ETH)
is true, 
the following results hold for a graph $G$ of $n$ nodes: 
%
\begin{enumerate}
\item[\emph{(\textbf{i})}]
there is no $O^\ast\big(2^{o(n)}\big)$-time algorithm for exactly computing $\sdim(G)$, and 
\item[\emph{(\textbf{ii})}]
if $\sdim(G)\leq k$ then 
there is no $O^\ast\big(n^{o(k)}\big)$-time algorithm for exactly computing $\sdim(G)$. 
\end{enumerate}
\end{theorem}

\subsection{Brief Remark on the Proof of Theorem~\ref{main-thm}}

Our proof uses Theorem~\ref{fact1} whose proof is implicit in~\cite{OP07}. However, it is not the case that 
Theorem~\ref{fact1} can be simply ``plugged in'' to get a proof of our inapproximability results.
Just because a problem can 
be written as a node cover problem (as in Fact~\ref{fact1})
does not necessarily mean that it has the same inapproximability property for node cover
since, for example, non-trivial special cases of node cover do admit efficient polynomial time solution.
To show inapproximability we need to reduce appropriate ``hard'' instances 
of the node cover problem to that of computing $\sdim(G)$ (\IE, a reduction in the opposite direction) and 
moreover such a polynomial-time reduction must be gap-preserving in an appropriate way 
(see~\cite[Section 10.1.3]{AL96} for
descriptions of gap-preserving reductions). 
For readers unfamiliar with gap-preserving reduction proof techniques, see the 
excellent survey by Arora and Lund in~\cite{AL96}.

\section{Proof of Theorem~\ref{main-thm}}

The minimum node cover (\MNC) problem for a graph is defined as follows: 

\medskip

\begin{tabular}{r p{3.5in}}
\toprule
\textbf{Instance}: & 
an undirected graph $G=(V,E)$.
\\
[4pt]
\textbf{Valid Solution}: &
a set of nodes $V'\subseteq V$ such that $V'\cap \{u,v\}\neq\emptyset$ for every edge $\{u,v\}\in E$.
\\
[4pt]
\textbf{Objective}: & 
\emph{minimize} $|V'|$.
\\
[4pt]
\textbf{Related notation}: & 
$\displaystyle\text{\MNC}(G)=\min_{\forall \, \{u,v\}\,\in E \colon V'\cap \{u,v\}\neq\emptyset} \Big\{ \,\, \big|\,V'\,\big|\,\,  \Big\}$.
\\
\bottomrule
\end{tabular}

\medskip
\noindent
Let $G=(V,E)$ denote the input graph of $n$ nodes. 
Let $\widehat{G}$ and 
$\widetilde{G}$ be two graphs obtained from $G$ in the following manner:

\begin{itemize}
\item
$\widehat{G}=(V,\widehat{E})$ where 
$\{u,v\}\in \widehat{E}$ if and only if 
$u\neq v\,\,$ \textbf{and} $\,\,u\!\stackrel{s}{\leftrightsquigarrow}\! v$ is a maximal shortest path in $G$.
%
\item
$\widetilde{G}=(\widetilde{V},\widetilde{E})$ where 
$\widetilde{V}$ and $\widetilde{E}$ are obtained as follows:
\begin{itemize}
\item
Let $u_1,u_2,\dots,u_{\kappa}$ be the nodes in $G$ such that, for every $u_i$ ($1\leq i\leq \kappa$), 
there is a node $v_i\neq u_i$ in $G$ with 
the property that {\em$\Nbr\left(u_i\right)=\Nbr\left(v_i\right)$}.

\item
Let $\overline{G}=(V,\overline{E})$ be the (edge) complement of $G$, \IE, $\{u,v\}\in \overline{E} \equiv \{u,v\}\notin E$.

\item
Then,
$\widetilde{V}=V \cup \left\{x_1,x_2,\dots,x_{\kappa},y\right\}$ where $x_1,x_2,\dots,x_{\kappa},y\notin V$, and 
$\widetilde{E}=\overline{E} \,\, \cup \,\, \left( \, \bigcup_{j=1}^{\,\kappa} \left\{\,\{x_j,u_j\}\,\right\} \,\, \right) \,\, \cup \,\, \left( \, \bigcup_{y'\,\in\, \widetilde{V}\,\setminus\, \{y\}}\, \Big\{ \,\big\{\,y',y\,\big\} \, \Big\} \,\, \right)$.
\end{itemize}
\end{itemize}
%
We recall the following result implicit in~\cite{OP07}.

\begin{theorem}{\rm\cite{OP07}}\label{fact1}~\\
\medskip
\noindent
{\bf (a)}
{\em $\sdim(G)=\MNC(\widehat{G})$}, and $V'\subseteq V$ is a valid solution of \SMD on $G$ if and only if $V'$ is a valid solution of
\MNC\ on $\widehat{G}$.

\medskip
\noindent
{\bf (b)}
{\em$\diam(\widetilde{G})=2$} and {\em$\sdim(\widetilde{G})=\kappa+\MNC(G)$}.
\end{theorem}

\noindent
A proof of Theorem~\ref{fact1} is implicit in~\cite{OP07}.
For reader's benefit, we provide a self-contained proof of
Theorem~\ref{fact1} in~\ref{ap1} using elementary graph theory.

\bigskip
\noindent
\textbf{Proof of Theorem~\ref{main-thm}(a)}

\smallskip
\noindent
Since 
$\sdim(G)=\MNC(\widehat{G})$, 
and 
both $G$ and $\widehat{G}$ 
have the same number of nodes, the claim follows by 
applying known algorithms for node cover on 
$\widehat{G}$. 
More precisely, 
\begin{itemize}
\item
the $2$-approximation follows from 
a well-known $2$-approximation algorithm for \MNC~\cite[Theorem 1.3]{V01},
\item
the $O^\ast\big(2^{\,0.287\,n}\big)$-time exact solution algorithm follows from the 
$O^\ast\big(2^{\,0.287\,n}\big)$-time exact algorithm for maximum independent 
set\footnote{Nodes \emph{not} in an independent set form a \emph{valid} solution of the node cover problem~\cite{GJ79}.} 
problem in~\cite{FGK09}, and 
\item
the $O\big(1.2738^k+n\,k\big)$-time exact computation algorithm 
follows from the 
$O\big(1.2738^k+n\,k\big)$-time exact algorithm 
for minimum node cover of 
$\widehat{G}$ 
provided $\MNC(\widehat{G})\leq k$~\cite{CKX10}.
\end{itemize}

\noindent
\textbf{Proof of Theorem~\ref{main-thm}(b)}

\smallskip
\noindent
Consider the standard Boolean satisfiability problem (\SAT)~\cite{GJ79} and let $\Phi$ be an input instance of \SAT. Our starting point 
is the following inapproximability result proved by Khot and Regev~\cite{KR08}:
\begin{quote}
\em
Assuming \UGC is true, there exists a polynomial time algorithm that transforms a given instance $\Phi$ of \SAT 
to an input instance graph $G=(V,E)$ of \MNC\ with $n$ nodes such that, for any constant $0<\eps<\frac{1}{4}$, the following holds:

\begin{tabular}{l l}
& 
\\
[-6pt]
($\star$) & 
\begin{tabular}{r l}
\textbf{(YES case)} & if $\Phi$ is satisfiable then $\MNC(G)\leq \left( \frac{1}{2}+\eps\right) n$, and 
\\
\\
[-10pt]
\textbf{(NO case)} & if $\Phi$ is not satisfiable then $\MNC(G)\geq \left( 1 - \eps\right) n$.
\end{tabular}
\end{tabular}
\end{quote}
Consider such an instance $G$ of \MNC\ as generated by the above transformation. 
Let $k=1+\left\lfloor \log_2 n \right\rfloor$ and let $b(j)=b_{k-1}(j)\,b_{k-2}(j)\dots\, b_1(j)\,b_0(j)$ be 
the binary representation of an integer $j\in\{1,2\dots,n\}$ using 
\emph{exactly} $k$ bits (\EG, if $n=5$ then $b(3)=\stackrel{b_2(3)}{0}\,\stackrel{b_1(3)}{1}\,\stackrel{b_0(3)}{1}\,\,$).
Let $u_1,u_2,\dots,u_n$ be an arbitrary ordering of the nodes in $V$.
We first construct the following graph $G^+=(V^+,E^+)$ from $G$:
\begin{itemize}
\item
$V^+=V \cup V_1^+$ where $V_1^+=\left\{ v_1,v_2,\dots,v_{k-1},y \right\}$ is a set of $k$ new nodes, and 
\item
$\displaystyle E^+=E \,\, \cup \,\, \left( \, \bigcup_{j=1}^{n} \, \left\{ \, \{ u_j,v_\ell \} \,\, | \,\, b_\ell(j)=1 \right\} \,\, \right) \,\, 
\cup \,\, \left( \, \bigcup_{j=1}^{k-1} \, \Big\{ \, \{ y,v_j \} \, \Big\} \,\right)$.
\end{itemize}
Thus $|V^+|=n+k$ and $|E^+|<|E|+\frac{n\,k}{2}+k$. Now, note that:
\begin{itemize}
\item
if $V'\subseteq V$ is a solution of \MNC\ on $G$, then $V' \cup V_1^+$ is a solution of \MNC\ on $G^+$, implying 
$\MNC(G^+)\leq\MNC(G)+k$, and conversely, 
\item
if $V'\subseteq V^+$ is a solution of \MNC\ on $G^+$, then $V' \setminus V_1^+$ is a solution of \MNC\ on $G$, 
implying $\MNC(G)\leq\MNC(G^+)$.
\end{itemize}
Combining the above inequalities with that in ($\star$), we have 

\begin{tabular}{l l}
& 
\\
[-6pt]
\hspace*{-0.35in}
($\star\star$) & 
\hspace*{-0.25in}
\begin{tabular}{r l}
\textbf{(YES case)} & if $\Phi$ is satisfiable then $\MNC(G^+)<\left( \frac{1}{2}+\eps\right) n+\log_2 n+1$, and 
\\
\\
[-10pt]
\textbf{(NO case)} & if $\Phi$ is not satisfiable then $\MNC(G^+)\geq \left( 1 - \eps\right) n$.
\end{tabular}
\\
\\
[-6pt]
\end{tabular}

\noindent
We now build the graph $\widetilde{G^+}=(\widetilde{V^+},\widetilde{E^+})$ from $G$ using the construction in 
Theorem~\ref{fact1}\textbf{(b)}.

\begin{claim}\label{fix-par}
No two nodes in $\widetilde{G^+}$ have the same neighborhood.
\end{claim}

\begin{proof}
The following careful case analysis proves the claim:
\begin{itemize}
\item
For any $i\neq j$, since $b(i)\neq b(j)$, there exists an index $t$ such that $b_t(i)\neq b_t(j)$, say 
$b_t(i)=0$ and $b_t(j)=1$. Thus, $\Nbr\left(u_i\right)\neq\Nbr(u_j)$ since 
$v_t\in\Nbr(u_j)$ but $v_t\notin\Nbr\left(u_i\right)$.
\item
Since $b(i)\neq 0$ for any $i$ and $b(1),b(2),\dots,b(n)$ are distinct binary numbers each of exactly $k$ bits, for any $t\neq t'$
there is an index $i$
such that $b_{t}(i)\neq b_{t'}(i)$, say $b_{t}(i)=0$ and $b_{t'}(i)=1$. Thus, $\Nbr\left(v_t\right)\neq\Nbr\left(v_{t'}\right)$ since 
$u_i\in\Nbr\left(v_{t'}\right)$ but $u_i\notin\Nbr\left(v_t\right)$.
\item
For any $i$ and $j$, $\Nbr\left(u_i\right)\neq\Nbr(v_j)$ since $y\in\Nbr(v_j)$ but $y\notin\Nbr\left(u_i\right)$.
\item
For any $i$, $b(i)\neq 0$ and thus there exists an index $j$ such that $b_j(i)=1$. 
This implies $u_j\in\Nbr\left(v_i\right)$ but $u_j\notin\Nbr(y)$ and therefore $\Nbr\left(v_i\right)\neq\Nbr(y)$.
\item
Since $G$ is a connected graph, for every node $u_i$ there exists a node $u_j$ such that $\left\{u_i,u_j\right\}\in E^+$.
Thus, $u_j\in\Nbr\left(u_i\right)$ but $u_j\notin\Nbr(y)$, implying $\Nbr\left(u_i\right)\neq\Nbr(y)$.
\end{itemize}
\end{proof}

By the above claim, $\kappa=0$ and
$\sdim(\widetilde{G^+})=\MNC(G^+)$ by Theorem~\ref{fact1}\textbf{(b)}.
Thus, setting $\eps'=\eps+\frac{\log_2 n + 1}{n}$ and noting that $\eps'$ can be any arbitrarily small constant
since $\eps$ is an arbitrarily small constant, it follows from $(\star\star)$ that 

\begin{tabular}{l l}
& 
\\
[-6pt]
\hspace*{-0.35in}
($\star\!\star\!\star$) & 
\hspace*{-0.3in}
\begin{tabular}{r l}
\textbf{(YES case)} & \hspace*{-0.15in} if $\Phi$ is satisfiable then $\sdim(\widetilde{G^+})=\MNC(G^+)<\left( \frac{1}{2}+\eps'\right) n$, and 
\\
\\
[-10pt]
\textbf{(NO case)} & \hspace*{-0.15in} if $\Phi$ is not satisfiable then $\sdim(\widetilde{G^+})=\MNC(G^+)\geq \left( 1 - \eps'\right) n$.
\end{tabular}
\\
\\
[-6pt]
\end{tabular}

\noindent
This proves Theorem~\ref{main-thm}\textbf{(b)(i)} since 
$\diam(\widetilde{G^+})=2$ by Theorem~\ref{fact1}\textbf{(b)}.

\medskip
To prove Theorem~\ref{main-thm}\textbf{(b)(ii)}, we modify the graph $\widetilde{G^+}$ to a new graph 
$G'=(V',E')$ by splitting every edge into a sequence of two edges, \IE, for every edge $\{u,v\}$ in  $\widetilde{G^+}$ we add a new node
$x_{u,v}$ in $G'$ and replace the edge $\{u,v\}$ by the two edges $\{u,x_{u,v}\}$ and $\{v,x_{u,v}\}$. Clearly $G'$ is bipartite
since all its cycles are of even length and $\diam(G')\leq 2\,\diam(\widetilde{G^+})=4$. 

\begin{claim}
$\sdim(\widetilde{G^+})
=
\MNC(\widehat{\widetilde{G^+}})
=
\MNC(\widehat{G'})
=
\sdim(G')
$.
\end{claim}

\begin{proof}
No maximal shortest path in $G'$ ends at 
a node $x_{u,v}$ for any distinct pair of nodes $u$ and $v$. 
Indeed, if a maximal shortest path $\cP$ from some node $z$ ends at some $x_{u,v}$, it must use one of the two edges 
$\{u,x_{u,v}\}$ and $\{v,x_{u,v}\}$, say $\{u,x_{u,v}\}$. Then adding the edge $\{v,x_{u,v}\}$ to the path $\cP$ 
provide a shortest path between $v$ and $z$, and thus $\cP$ was not maximal. 
Using this and the construction in Theorem~\ref{fact1}\textbf{(a)}, we have  
$\widehat{\widetilde{G^+}}=\widehat{G'}$ and therefore
$
\sdim(\widetilde{G^+})
=
\MNC(\widehat{\widetilde{G^+}})
=
\MNC(\widehat{G'})
=
\sdim(G')
$.
\end{proof}

As a result, the inapproximability result for $\sdim(\widetilde{G^+})$ directly translates to that for $\sdim(G')$, 
and concludes the proof.

\bigskip
\noindent
\textbf{Proof of Theorem~\ref{main-thm}(c)}

\smallskip
\noindent
The same proof as in {\bf (b)} works provided, instead of the result in~\cite{KR08},
our starting point is the following result shown by Dinur and Safra~\cite{DS05}\footnote{\comment{\color{blue}}Note that 
$\left( { \frac{71-31\sqrt{5}}{2} } \right) \,/\, \left( { \frac{\sqrt{5}-1}{2} } \right) = 10\sqrt{5}-21$.}:
\begin{quote}
\em
Assuming P$\neq\!\!\!\NP$, there exists a polynomial time algorithm that transforms a given instance $\Phi$ of \SAT 
to an input instance graph $G=(V,E)$ of \MNC\ with $n$ nodes such that, for any constant $0<\eps<16-8\sqrt{5}$ and 
for some $0<\alpha<2n$, the following holds:

\begin{tabular}{l l}
& 
\\
[-6pt]
($\star$) & 
\begin{tabular}{r l}
\textbf{(YES case)} & if $\Phi$ is satisfiable then $\MNC(G)\leq \left(\frac{\sqrt{5}-1}{2}+\eps\right)\alpha$, and 
\\
\\
[-10pt]
\textbf{(NO case)} & if $\Phi$ is not satisfiable then $\MNC(G)\geq \left( \frac{71-31\sqrt{5}}{2}-\eps\right)\alpha$.
\end{tabular}
\end{tabular}
\end{quote}

\bigskip
\noindent
\textbf{Proof of Theorem~\ref{main-thm}(d)}
We first show how to prove Theorem~\ref{main-thm}\textbf{(d)(i)}.
Suppose, for the sake of contradiction, that there does exist a 
$O^\ast\big(2^{o(n)}\big)$-time algorithm that exactly computes $\sdim(G)$.
We start with an instance $\Phi$ of \tSAT having $n$ variables and $m$ clauses.
The ``sparsification lemma'' in~\cite{IPZ01} proves the following result: 
\begin{quote}
\em
for every constant $\eps>0$, there is a constant $c>0$ such that 
there exists a $O\big(2^{\,\eps n}\big)$-time algorithm that produces 
from $\Phi$ 
a set of $t$ instances $\Phi_1,\dots,\Phi_t$ of \tSAT on these $n$ variables 
with the following properties:
\begin{itemize}
\item
$t\leq 2^{\,\eps n}$,
\item
each $\Phi_j$ is an instance of \tSAT with $n_j\leq n$ variables and $m_j\leq cn$ clauses, and 
\item
$\Phi$ is satisfiable if and only if at least one of 
$\Phi_1,\dots,\Phi_t$ is satisfiable.
\end{itemize}
\end{quote}
For each such above-produced \tSAT instance $\Phi_j$, 
we now use the ``classical textbook'' reduction from \tSAT to the node cover problem 
(\EG, see~\cite[page 54]{GJ79})
producing an instance 
$G=(V,E)$ of \MNC\ of $|V|=3n_j+2m_j\leq (3+2c)\,n$ nodes and 
$|E|=n_j+m_j\leq (1+c)\,n$ edges
such that $\Phi_j$ is satisfiable if and only if 
$\MNC(G)=n_j+2m_j$.
Moreover, it is also easy to check that this classical reduction does \emph{not} 
produce two nodes in $V$ that have the \emph{same} neighborhood.
Thus, setting $\kappa=0$
in Theorem~\ref{fact1}\textbf{(b)} we get 
$\sdim(\widetilde{G})=\MNC(G)$ where  
$\widetilde{G}$ is a graph with 
$\widetilde{n}=|\widetilde{V}|=|V|+1\leq (3+2c)\,n + 1$ nodes.
By assumption, we can compute 
$\sdim(\widetilde{G})$ 
in $O^\ast\big(2^{o(\,\widetilde{n}\,)}\big)$ time, and 
and consequently $\MNC(G)$
in $O^\ast\big(2^{o(n)}\big)$ time, 
which leads us to decide in 
$O^\ast\big(2^{o(n)}\big)$ time 
if $\Phi_j$ is satisfiable.
Since 
$t\leq 2^{\,\eps n}$
for any constant $\eps>0$,
this provides a 
$O^\ast\big(2^{o(n)}\big)$-time 
algorithm for \tSAT, contradicting \ETH.

To prove 
Theorem~\ref{main-thm}\textbf{(d)(ii)}
suppose again, for the sake of contradiction, that 
there exists a $O^\ast\big(n^{o(k)}\big)$-time algorithm for exactly computing $\sdim(G)$
if $\sdim(G)\leq k$.
Our proof is very similar to the previous one, but this time 
we start with the following lower bound result on parameterized complexity (\EG, see~\cite[Theorem 14.21]{CygFKLMPPS2015}): 
\begin{quote}
{\em 
assuming \ETH to be true, if $\MNC(G)\leq k$
then there is \emph{no} $O^\ast\big(n^{o(k)}\big)$-time algorithm for exactly computing $\MNC(G)$.
}
\end{quote}
Using the encoding as described in part \textbf{(b)} of this proof with the corresponding Claim~\ref{fix-par}, we can 
set $\kappa=0$ in 
Theorem~\ref{fact1}\textbf{(b)}
to obtain the graph 
$
\widetilde{G^+}
=
(\widetilde{V^+},\widetilde{E^+})
$
such that 
$
\widetilde{n^+}
=
| \widetilde{V^+} | 
= 
|V| + \big(1+\left\lfloor \log_2 n \right\rfloor \big) + 1
= 
n +  \left\lfloor \log_2 n \right\rfloor + 2
$
and 
$\sdim(\widetilde{G^+})=\MNC(G)$.
By our assumption, 
we can 
compute $\sdim(\widetilde{G^+})$ in 
$O^\ast\Big( ( {\widetilde{n^+} ) }^{o(k)}\Big)$-time algorithm if $\sdim(G)\leq k$.
This then provides an  algorithm running in 
$O^\ast\Big( ( {\widetilde{n^+} ) }^{o(k)}\Big)=O^\ast\big( n^{\,o(k)}\big)$ 
time if 
$\MNC(G)=\sdim(G) \leq k$, contradicting \ETH.

\section{Conclusion}

In this communication we have shown that the worst-case computational complexity for computing the strong metric
dimension for many graphs behaves in a manner similar to the minimum node cover problem. 
However, several interesting computational complexity questions still remain open, such as the following.
\begin{itemize}
\item
Does the $(2-\eps)$-inapproximability result for computing $\sdim(G)$ hold even when $G$ is bipartite and $\diam(G)\leq 3$ ?
\item
Are there interesting non-trivial classes of graphs for which $\sdim(G)$ can be computed in polynomial time ?
\item
In the context of kernelization for parameterized algorithms (\EG, see~\cite{CygFKLMPPS2015}), is there a 
linear kernel for \SMD ? 
\end{itemize}

\appendix

\section{Proof of Theorem~\ref{fact1}}
\label{ap1}

A proof of Theorem~\ref{fact1} is implicit in~\cite{OP07}.
For the benefit of the reader, we provide a self-contained proof of
Theorem~\ref{fact1} here using elementary graph theory.

\medskip
\noindent
{\bf (a)}
Let $u\!\stackrel{s}{\leftrightsquigarrow}\! v$ be a maximal shortest path in $G$.
Suppose that we select neither $u$ nor $v$ in a solution of solution of \SMD on $G$.
Then there exists no node $x$ in our solution of \SMD on $G$ such that 
$x\blacktriangleright \{u,v\}$, implying our solution of \SMD on $G$ is \emph{not} a valid solution and thereby
showing $\sdim(G)\geq\MNC(\widehat{G})$. To prove $\sdim(G)\leq\MNC(\widehat{G})$, 
suppose that we select at least one end-point of every maximal shortest path 
in $G$. Consider any pair of nodes $u$ and $v$. If at least one of $u$ or $v$, say $u$, is selected in a 
solution of \SMD on $G$, 
then 
$u\blacktriangleright \{u,v\}$.
Otherwise, 
$u\!\stackrel{s}{\leftrightsquigarrow}\! v$ 
is {\em not} a maximal shortest path,
and let 
$x\!\stackrel{s}{\leftrightsquigarrow}\! y$ 
be a maximal shortest path containing $u$ and $v$. Then, we have selected at least one of $x$ or $y$, say $x$, 
in a solution of \SMD on $G$, and 
$x\blacktriangleright \{u,v\}$.

\medskip
\noindent
{\bf (b)}
It follows from the construction of 
$\widetilde{G}$
that 
$\diam(\widetilde{G})=2$
since any pair of nodes has a shortest path of length at most $2$ between them via $y$.
Note that, for any pair of nodes $u$ and $v$, 
$\Nbr\left(u\right)=\Nbr\left(v\right)$ in $G$ 
if and only if 
$\Nbr\left(u\right)=\Nbr\left(v\right)$ in $\overline{G}$.
To show 
$\sdim(\widetilde{G})\leq\kappa+\MNC(G)$, 
let $S\subset V$ be the set of nodes in a minimum node cover of $G$ of cardinality 
$\MNC(G)$.
Consider the set of $\kappa+\MNC(G)$ nodes in 
$S'=S \cup \left\{x_1,x_2,\dots,x_{\kappa}\right\}$
as a possible 
solution of \SMD on $\widetilde{G}$.
To show that this is indeed a valid solution, consider any pair of nodes $u$ and $v$ in 
$\widetilde{G}$.
Then the following simple case analysis suffices:
\begin{itemize}
\item
Suppose that at least one of $u$ and $v$ is $x_i$ for some $i$. Then,  
$S'\ni x_i\blacktriangleright \{u,v\}$.
\item
Otherwise, 
suppose that one of $u$ and $v$, say $u$, is $y$ (and thus $v\in V$). Select a node $x_i\in S'$ 
such that $\left\{x_i,v\right\}\notin \widetilde{E}$. 
Then the shortest path of length $2$
from $x_i$ to $v$ formed by the edges 
$\left\{x_i,y\right\}$ and 
$\left\{y,v\right\}$ shows that 
$S'\ni x_i\blacktriangleright \{u,v\}$.
\item
Otherwise, 
if 
$\{u,v\}\in E$ 
then at least one of $u$ and $v$, say $u$, is in $S'$ and
$u\blacktriangleright \{u,v\}$.
\item
Otherwise, $\{u,v\}\notin E$. 
Thus, $\{u,v\}\in \widetilde{E}$. 
If at least one of $u$ and $v$, say $u$, is in $S$ then 
$u\in S'$ and 
$u\blacktriangleright \{u,v\}$.
Otherwise, both of $u$ and $v$ are not in $S$, and 
there are the following two sub-cases to consider.
\begin{itemize}
\item
At least one of $u$ and $v$, say $u$, is $u_i$ for some $i$.
Then the shortest path of length $2$
from $x_i$ to $v$ formed by the edges 
$\left\{x_i,u_i\right\}$ and 
$\left\{u_i,v\right\}$ shows that 
$S'\ni x_i\blacktriangleright \{u,v\}$.
\item
Otherwise,
$\Nbr\left(u\right)\neq\Nbr\left(v\right)$ in $G$, which 
implies that there exists a node $u'\in V$ such that 
$u'$ is adjacent to exactly one of $u$ and $v$, say $u$.
Thus, 
$\left\{u,u'\right\}\notin \widetilde{E}$ but
$\left\{v,u'\right\}\in \widetilde{E}$.
Note that $u\notin S$ and $\left\{u,u'\right\}\in E$ implies 
$u'$ is in $S$.
Then the shortest path of length $2$
from $u'$ to $u$ formed by the edges 
$\left\{u',v\right\}$ and 
$\left\{v,u\right\}$ shows that 
$S'\ni u'\blacktriangleright \{u,v\}$.
\end{itemize}
\end{itemize}
To show 
$\sdim(\widetilde{G})\geq\kappa+\MNC(G)$, 
let $S'\subset \widetilde{V}$ be the set of 
$\sdim(\widetilde{G})$
nodes in an optimal 
solution of \SMD on $\widetilde{G}$.
Consider the set of nodes in 
$S=S' \setminus \left\{x_1,x_2,\dots,x_{\kappa},y\right\}$
as a possible solution of the 
node cover problem of $G$.
We first show that $S$ is in fact a valid node cover of $G$.
Since 
$\diam(\widetilde{G})=2$, 
any shortest path in $G$ is of length at most $2$.
Consider an edge $\{u,v\}\in E$ and suppose that both $u$ and $v$ are not in $S$ (and thus also not in $S'$).
Since $\{u,v\}\notin\widetilde{E}$, the length of any shortest path between $u$ and $v$ is exactly $2$, and 
thus no node $x\in\widetilde{V}\setminus\{u,v\}$ can strongly resolve the pair of nodes $u$ and $v$, 
resulting in a contradiction that $S'$ is a solution of \SMD on $\widetilde{G}$.
Thus, $S$ is a node cover of $G$ and $\MNC(G)\leq |S|$.
To show that $|S|=|S'|-\kappa$, note that:
\begin{itemize}
\item
Every $x_i$ must belong to $S'$ since otherwise no node in $S'$ can strongly resolve the pair of nodes $x_i$ and
$x_j$ for any $j\neq i$.
\item
Since every $x_i$ belongs to $S'$, the node $y$ does not need to belong to $S'$.
\end{itemize}

\end{document}